\documentclass[utf8]{FrontiersinHarvard} 

\usepackage{url,hyperref,lineno,microtype,subcaption}
\usepackage[onehalfspacing]{setspace}
\usepackage{xcolor}

\usepackage{soul}
\usepackage{pdfpages}

\def\keyFont{\fontsize{8}{11}\helveticabold }
\def\firstAuthorLast{Xu {et~al.}} 
\def\Authors{Zigong\,Xu\,$^{1,*}$, Jingnan Guo\,$^{2,3}$ ,  Robert F.\,Wimmer-Schweingruber\, $^{1,4,*}$ , Mikhail I.\,Dobynde\,$^{2}$,  Patrick K\"uhl\,$^{1}$ , Salman Khaksarighiri\,$^{1}$ and  Shenyi Zhang\,$^{4,5,6}$}
\def\Address{$^{1}$ Institute of Experimental and Applied Physics, Kiel University, Kiel, Germany \\
$^{2}$ School of Earth and Space Sciences, University of Science and Technology of China, Hefei, PR China\\
$^{3}$ CAS Center for Excellence in Comparative Planetology, Hefei, PR China\\
$^{4}$ National Space Science Center, Beijing, PR China \\
$^{5}$Beijing Key Laboratory of Space Environment Exploration, Beijing, PR China\\
$^{6}$University of Chinese Academy of Science, Beijing, PR China}
\def\corrAuthor{Zigong Xu \\ Institute of Experimental and Applied Physics, Kiel University, 24118 Kiel, Germany}

\def\corrEmail{xu@physik.uni-kiel.de}

\begin{document}
\onecolumn
\firstpage{1}

\title[LND proton measurements]{Primary and albedo protons detected by the Lunar Lander Neutron and Dosimetry (LND) experiment on the lunar farside} 

\author[\firstAuthorLast ]{\Authors} 
\address{} 
\correspondance{} 

\extraAuth{Robert F.\,Wimmer-Schweingruber \\  Institute of Experimental and Applied Physics, Kiel University, 24118 Kiel, Germany, \\ National Space Science Center, Beijing, China,  wimmer@physik.uni-kiel.de}

\maketitle

\begin{abstract}

\section{}
The Lunar Lander Neutron and Dosimetry (LND) Experiment aboard the Chang'E-4 Lander on the lunar far-side measures energetic charged and neutral particles and monitors the corresponding radiation levels. During solar quiet times, galactic cosmic rays (GCRs) are the dominating component of charged particles on the lunar surface. Moreover, the interaction of GCRs with the lunar regolith also results in upward-directed albedo protons which are measured by the LND. In this work, we used calibrated LND data to study the GCR primary and albedo protons. We calculate the averaged GCR proton spectrum in the range of 9-368 MeV and the averaged albedo proton flux between 64.7 and 76.7 MeV from June 2019 (the 7th lunar day after Chang'E-4's landing) to July 2020 (the 20th lunar day). We compare the primary proton measurements of LND with the Electron Proton Helium INstrument (EPHIN) on SOHO. The comparison shows a reasonable agreement of the GCR proton spectra among different instruments and illustrates the capability of LND. Likewise, the albedo proton measurements of LND are also comparable with measurements by the Cosmic Ray Telescope for the Effects of Radiation (CRaTER) during solar minimum. Our measurements confirm predictions from the Radiation Environment and Dose at the Moon (REDMoon) model. Finally, we provide the ratio of albedo protons to primary protons for measurements in the energy range of 64.7-76.7 MeV which confirm simulations over a broader energy range.

\tiny
 \keyFont{ \section{Keywords:} LND, Moon, Lunar radiation environment, Galactic cosmic rays, Lunar albedo protons, Data calibration, Instrumentation} 
\end{abstract}

\section{Introduction}

The charged particle radiation environment on the lunar surface consists of Galactic Cosmic Rays (GCR), a small contribution from Anomalous Cosmic Rays (ACR), and a highly variable, sporadic contribution from Solar Energetic Particles (SEPs). In addition, secondary albedo particles are created primarily by the GCR interaction with the lunar regolith \citep{TREIMAN1953PhRv, Dorman2004ASSL, Wilson2012}. In this work we focus on measurements by the Lunar Lander Neutron and Dosimetry experiment \citep[LND,][]{Wimmer2020SSR} of GCR primary and secondary (albedo) protons and compare them with state-of-art model predictions \citep{DobyndeREDMOON2021}. The space radiation measured by LND is a key concern for
human space flight and may pose limits on long-term crewed missions to the Moon or Mars \citep{cucinotta-etal-2011}.

The interaction of high-energy particles with the lunar regolith results in the production of secondary particles. Some of these particles can escape from the soil and can be measured as albedo particles. Obviously, they will also contribute to the radiation exposure of astronauts on the lunar surface, but they also provide information about conditions beneath the lunar surface.

\citet{Wilson2012} first distinguished protons measured by the Cosmic Ray Telescope for the Effects of Radiation \citep[CRaTER,]{spence2010} instrument from different directions and constructed the first albedo proton yield map between 60 and 150 MeV. They found that the ratio of albedo to primary protons is uniformly distributed over the lunar surface within $\sim$ 10\% uncertainty and that the average ratio of upward to downward particle during the minimum between solar cycles 23 and 24 (in 2009-2011) was about 0.38 $\pm$ 0.02. Simulations by \citet{Spence2013} show that such upward albedo particles can contribute a significant amount ($\sim$ 8.62\%) to the radiation dose with albedo protons accounting for 3.1\% of the total dose rate. Moreover, albedo neutrons have been used to detect and determine the subsurface hydrogen content \citep{Mitrofanov2016}. Likewise, \citet{Schwadron2016} used the CRaTER instrument to detect albedo protons and provided evidence for the existence of hydrated material in the lunar regolith at the polar regions, based on small variations of the proton yield. A follow-up study by \citet{Schwadron2017} analyzed the difference of the lunar albedo protons yield for the lunar sunrise and sunset terminators.

\citep{Looper2013} presented the CRaTER measurements of radiation environment near the lunar surface including GCRs and albedo particles. They also utilized a GEANT4 simulation to model the energy distribution of albedo particles, and the response of different detectors to various particle species that reach the instrument on the 50km height lunar orbit.

More recently, \citet{DobyndeREDMOON2021} developed the Radiation Environment and Dose at the Moon (REDMoon) model that describes the detailed radiation environment both on the lunar surface and in the lunar soil. In this model, the lunar body is described by concentric spherical layers of different densities and soil composition is based on Apollo 17 drill core results \citep{mckinney-etal-2006}.
The model is based on the GEANT4 Monte-Carlo particle transport code \citep{Agostinelli2003} and the FTFP$\_$BERT$\_$HP physics list was used to calculate the GCR-induced radiation environment on and below the lunar surface. Primary GCRs have an isotropic direction (in the upper half-sphere) and the output particle types and energy at each soil depth or zenith angle are recorded. Results from different solar modulation conditions are also derived.

\begin{figure}[!htp]
\centering
\includegraphics[width = 8cm, height =10cm]{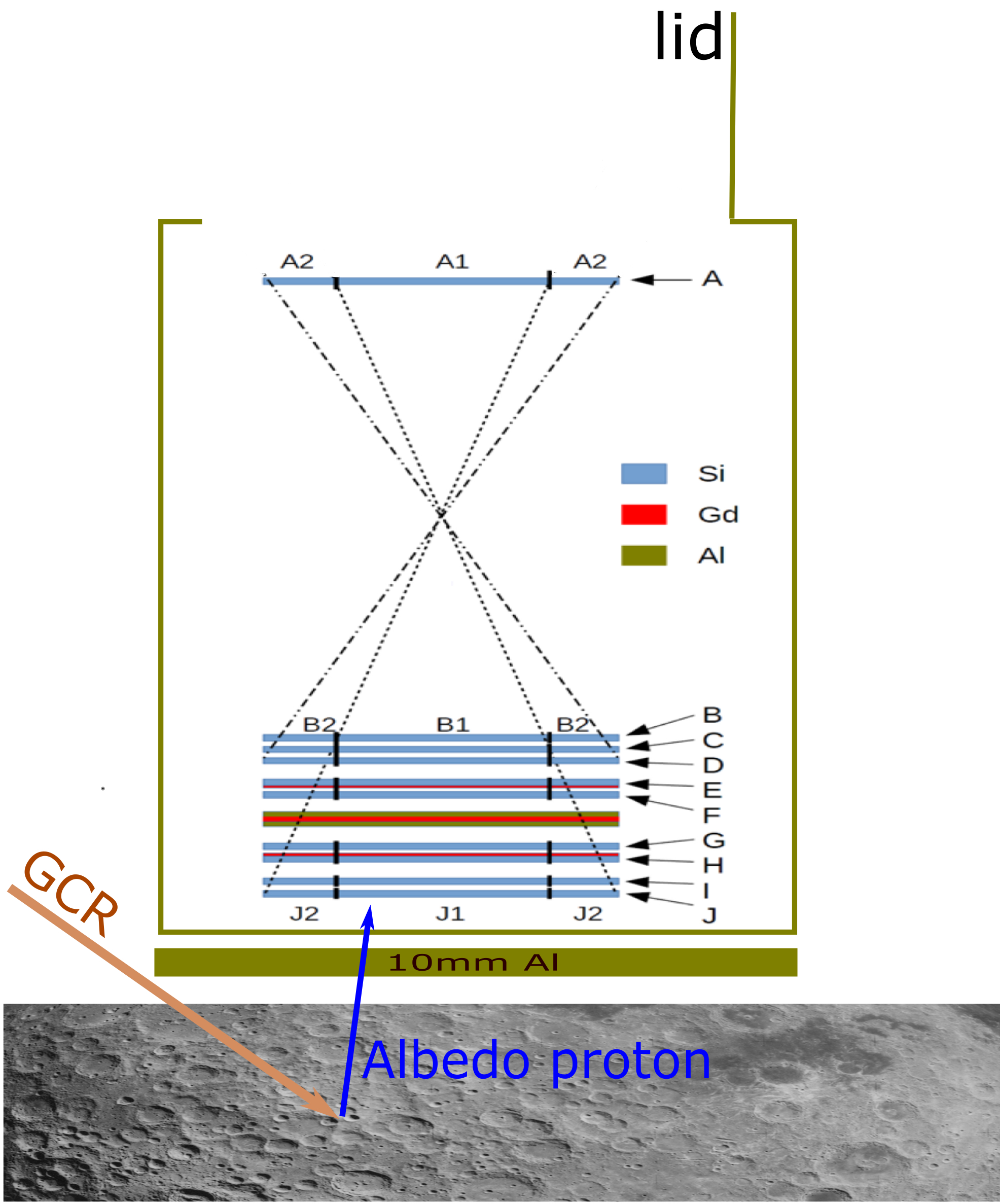}
\caption{Schematic view of the LND sensor head located above the lunar regolith with a 10-mm Al equivalent shielding (olive) between LND and the lunar soil. The LND sensor head consists of 10 segmented 500$\mu$m-thick Si detectors A -- J. The Gd (red) and Al (olive) absorbers are designed to detect uncharged particles. The 10-mm Al sheet is added to mimic the shielding by the material between the sensor head and the lunar soil. On the lunar surface, the interaction of high-energy GCR particles (mainly protons) with the soil generates secondary, upward-moving protons. A more detailed description of LND is given in \cite{Wimmer2020SSR}.}
\label{Fig:LND-structure}
\end{figure}

On January 3, 2019, China successfully landed the Chang'E-4 spacecraft on the far side of the Moon inside the von K\`arm\`an Crater. As one of its scientific payloads, Lunar Lander Neutron and Dosimetry experiment \citep[LND,][]{Wimmer2020SSR} aboard the Chang'E-4 Lander is a small instrument with the primary objective of monitoring the radiation level on the lunar surface. LND consists of ten 500 $\mu$m silicon detectors (labeled A-J) that are assembled as shown in Fig.~\ref{Fig:LND-structure}. 
The primary data products of LND include dynamic Total Ionizing Dose rate (TID) and Linear Energy Transfer (LET) spectra. The radiation dose rate is the radiation energy deposited by incoming radiation per unit time and unit mass of the absorber and is often measured in silicon detectors. LET is the energy that an ionizing particle transfers to the material per unit path length. Knowledge of both of these quantities is crucial in preparing  for human spaceflight and human exploration of the Moon.
Apart from these dosimetric quantities, LND also measures primary charged-particle energy spectra. Using the energy loss in its individual detectors, LND measures the primary energy of particles in the energy range between $\sim$10 MeV/nuc and a few hundred MeV/nuc and distinguishes different particle species, including electrons, protons, $^4$He, its isotope $^3$He, and heavy ions such as carbon, oxygen, nitrogen, and iron. The post-launch performance of LND and its capability of measuring protons have been partially verified through analysis of a weak and impulsive SEP event in May 2019 \citep{Xu2020ApJ}, with proton energy up to $\sim$20 MeV. Moreover, LND measures upward-directed albedo particles using the same method as was successfully applied using data from the Radiation Assessment Detector\citep[RAD,][]{Hassler2012SSRv} on Mars. \citet{2018Appel} determined the flux of albedo particles in the energy range 100-200 MeV on the Martian surface using a two-dimensional count density histogram. Here we use a similar method to distinguish upward proton fluxes from downward ones. 
. 

\citet{zhang-etal-2020} displayed the first measurement of TID and LET spectra during the first two lunar days after LND was switched on. During the solar minimum of Solar Cycle (SC) 24/25, the average total absorbed dose rate on the lunar surface reached 13.2 $\pm$ 1 $\mu Gy/h$. This value is consistent with the radiation dose that is measured by CRaTER in orbit after the conversion to the surface 
in January 2019. A comparison with the radiation level on the Mars surface shows that the dose rate on the lunar surface is about 15\% higher than that on Mars when GCR intensity reached a maximum during the deep solar minimum. Furthermore, the lunar surface dose rate is about two times higher than that measured in the International Space Station (ISS) by the 3D-DOSTEL instrument \citep{Berger2020JSWSC}.

More than three years into the Chang'E 4 mission, most of LND's detectors are still in good condition and operating well despite the grueling temperature differences between lunar day and night. 
Nevertheless, since a mishap on the dawn of the third lunar day some segments of four of the 10 detectors in LND have begun to suffer from increased noise levels. We believe that a premature opening of LND's lid led to a severe drop in the temperature of LND's sensor head. The mounting of LND's detectors led to unintended thermo-mechanical stress on the front detector (A) and the three detectors at the end of the particle telescope (H, I, and J). An adjustment of the thresholds of the affected detector segments has led to some changes in the LND data products compared to those presented in \cite{Wimmer2020SSR}. The changes are described in the appendix. 

In this work, we provide a detailed calibration of the GCR proton spectra measured by LND and present the first measurements of the albedo protons by LND. The paper is organized as follows:
In Section \ref{sec:Instrument}, we briefly describe LND, its measurement principles for primary and albedo protons, data calibration, the processing of LND data, and the simulation set-up used for interpreting the data. 
In Section \ref{sec:Measurement}, we present the LND measurements and compare processed LND data with other available data sets, i.e., the primary GCR measurements by the Electron Proton Helium INstrument \cite[EPHIN,][]{muller1995costep} on-board SOHO and the albedo protons measured by CRaTER onboard LRO \citep{Wilson2012,Schwadron2016}. We compare these measurements with the numerical cosmic ray model of Cosmic Ray Effects on Micro-Electronics Code \citep[ CREME96,]{Tylka1997}, Badhwar-O’Neill 2014 model \cite[BON14,][]{o2015badhwar} and the lunar radiation model REDMoon as mentioned above.
Section \ref{sec:conclusions} gives a summary and discussion. Finally, the appendix gives a detailed description of the current LND configuration and calibration parameters.

\section{Preparation of LND data}
\label{sec:Instrument}

\subsection{Instrument and the simulation set-up}
\label{sec:simulation-set-up}
A schematic structure of the LND sensor head is given in Fig.~\ref{Fig:LND-structure}. It consists of ten silicon detectors, each with a nominal thickness of 500~$\mu$m. Detectors are arranged in a charged-particle telescope configuration and labeled A through J from top to bottom. Each detector is segmented into an inner and an outer segment (labeled 1 \& 2, respectively), each with approximately the same area. LND uses coincidence measurements to detect the charged particles which requires particles to pass through the uppermost detector A and at least trigger the next detector, B.
In the middle of the sensor head, Al and Gd absorbers (shown in olive and red  in Fig.~\ref{Fig:LND-structure}) are used to detect the upward and downward flux of thermal neutrons (Gd has a very large cross-section for thermal neutrons). These absorbers increase the scattering of particles inside the sensor head and absorb part of the particle energy without providing a measurement thereof.  

Above detector A, the front window is covered by a lid that opens when LND operates, allowing particles to arrive at the detector without any obstacles. During lunar nights, the lander closes the lid to keep the sensor head and detectors warm enough for them to survive the cold lunar night.
The payload compartment of the Chang'E-4 lander which contains other scientific instruments is located beneath the LND sensor head and thus provides extra shielding for upward albedo particles before they reach LND. However, the details of that shielding are unknown and an estimation of 10-mm Al shielding equivalent was provided by the Chang'E-4 team. We thus add a 10-mm Al sheet between LND and the lunar soil in our model for estimating the albedo contribution to LND as shown in Fig.~\ref{Fig:LND-structure}.
Consequently, the shielding material raises the lowest energy that albedo protons are required to have in order to arrive at the LND detector. A more detailed description of LND is given in \cite{Wimmer2020SSR}.

We employed the GEANT4 toolkit \citep{Agostinelli2003}, version 10.4.1., using the QGSP\_BERT physics list to simulate the detector response to different particles from different directions at the relevant energies.

To simulate the particles from above, we placed a square-shaped planar source right above the front detector A with a size that is larger than the size of detector A and fully covers the field of view (FOV) of LND, i.e., the inner dashed line of the combination of A1 and B1 in Fig.~\ref{Fig:LND-structure}. Similarly, a square particle source is placed below the 10mm-Al sheet when simulating upward particles. The sources are isotropic and the energy spectrum is a power law with index -1, i.e., logarithmically flat, to allow easy scaling to any other spectral index \footnote{The “logarithmically flat” spectrum appears “flat” or horizontal when plotted in a log-log plot. Thus, bins are evenly spaced in a logarithmic scale and have the same number of counts per bin.}, as discussed, e.g., in \cite{Guo2019JSWSC}, but is a standard process. The spectrum is in differential flux units. We then obtain the averaged geometric factors from such a simulation and use them to convert detected counts to flux. 

\begin{figure}[!ht]
\centering
\includegraphics[width=\textwidth,]{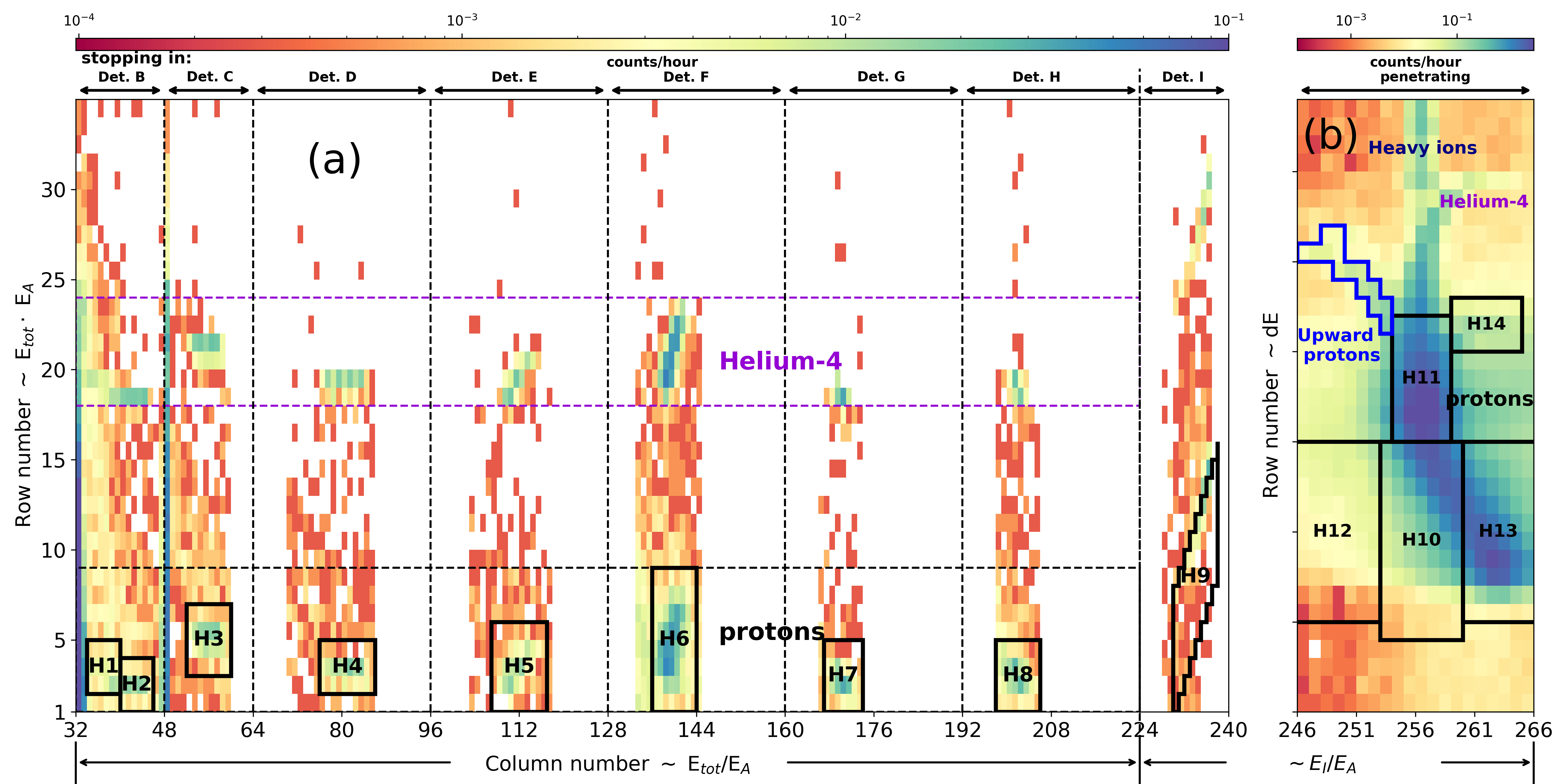}
\caption{The (a) panel displays the "Xmas plot" of part of the charged particles stopping between detectors B and I. The empty area represents the part of the LND memory that is not transmitted to the ground. Panel (b) is the Xmas plot between columns 246 and 266. The primary proton channels and albedo proton channel are defined accordingly in those two panels, marked by black and blue boxes. The quantities used to calculate the entry "pixel" in the Xmas plot are calculated according to the quantities summarized in Tab.~5 in \cite{Wimmer2020SSR}.}
\label{Fig:2Dhistogram}
\end{figure}

\subsection{Xmas plot}
\label{sec:xmasplot}

The Xmas plot\footnote{We call this plot ``Xmas plot" because our group in Kiel used a similar plot as our 2017 Xmas card.} is the primary data product of LND that is generated onboard and represents the memory space of the LND flight model. All LND measurements are stored in a 274 $\times$ 64 matrix though not all elements of this matrix are transmitted to Earth. The memory space refreshes every hour; hence the time cadence of the Xmas plot is 1 hour.
The Xmas plot can be divided up column-wise into the following regions (from columns 0 to 274): thermal neutrons, fast neutrons, TID, LET, and charged particles. More details of the data products can be found in \citet[Sec.~4]{Wimmer2020SSR}.
In this section, we focus on charged particles and explain the generation of the corresponding two-dimensional density histogram in the Xmas plot which is important to understand the calibration of LND data. Part of the charged particles measurement is displayed in panels (a) and (b) of Fig.~\ref{Fig:2Dhistogram} 

The charged particles in the Xmas plot can be further divided up into two parts, \textit{stopping} and \textit{penetrating} particles. The stopping particles have energies between $\sim$ 8 and $\sim$ 35 MeV/nuc. LND thus measures SEPs, ACRs, and lower energy GCRs in this energy range. When arriving at LND on the lunar surface, particles within this energy range can first trigger and penetrate detector A, then pass the following detectors, depositing (part of) their kinetic energy. Depending on their primary energy, some of these particles will stop in one of the detectors between B and I (as indicated by B, C, etc.\,along the top of Fig.~\ref{Fig:2Dhistogram}(a)) and hence deposit all of their energy. The higher their primary energy, the deeper they get in the detector stack. The upper limit is $\sim$35 MeV/nuc  for protons and helium nuclei stopping in detector I, but different for heavier ions \citep{Wimmer2020SSR}. A small uncertainty in their primary energy remains due to noise in the individual detectors.
Particles with energies above 35 MeV/nuc will penetrate detector I and stop in detector J or penetrate it, i.e., all detectors from A to J are triggered. Penetrating particles may come from above or below. 
Panels (a) and (b) of Fig. \ref{Fig:2Dhistogram} show those parts of the Xmas plot that belong to stopping and penetrating particles. A precise description of the quantities used to populate the Xmas plot are given in Tab. 5 of \citet{Wimmer2020SSR}, but a short summary is given here for convenience. For particles stopping between detectors B and H, the values along the x axis are a monotonous function of the ratio E$_{\rm total}$/E$_{A}$. The y axis is ordered according to the product E$_{total}\cdot$E$_{A}$. The ratio and product are mapped to row and column values internally in the instrument. Thus, the column regions marked by B, C, D are similar to traditional d$E$/d$x$ - E plots, but rotated and compressed into d$E$/d$x$ $\cdot$ $E_{\rm total}$ vs. d$E$/d$x$ / $E_{\rm total}$ space. In this space the "1/E" energy loss appears as a horizontal line and the penetrating particles do not show up because they are detected in the next detector. The horizontal (energy) axis is compressed by the division by E$_{\rm A}$. Particles stopping in detectors E, F, etc.\,are treated somewhat differently, see \citet{Wimmer2020SSR} for more details. The quantities used for penetrating particles and particles stopping in detector I are similar. The x axis is ordered by $E_I/E_A$ and the y axis is ordered by the sum of energy depositions in detectors B, C, and D (see sec.~\ref{sec:albedo-response} for the exact definition of penetrating particles).
Stopping particles populate columns 32 -- 240, while penetrating particles are between 246 and 266. The data in the empty regions in Fig.~\ref{Fig:2Dhistogram} (a) are not transmitted back to Earth. 

The proton channels that are used in this work are outlined by the black boxes in panels (a) and (b) of Fig.~\ref{Fig:2Dhistogram} and further indicated by H1 - H14. We call them DPS boxes, where DPS stands for ``data product scheduler". Count numbers inside DPS boxes are read out from memory every minute. Therefore the cadence of primary proton data is 1 minute. These high-time-resolution proton data are important during the onsets of SEP events \citep{Xu2020ApJ}. The pixels with high count rates around row 20 in Fig.~\ref{Fig:2Dhistogram}(a) are $^4$He counts. Heavier ions would lie at even higher rows.

One of the advantages of the Xmas plot data product is that one can define one's own data product. For this work, we defined a mask for upward-moving protons. It is marked by the blue box in panel (b) of Fig.\ref{Fig:2Dhistogram}. The positions of those pixels were determined manually based on simulation results. Because such particles are rare, the one-hour time resolution of the Xmas plot is more than sufficient. In the following sections, we explain this data product in depth.

The Xmas plot in Fig.~\ref{Fig:2Dhistogram} was accumulated between the 7th and the 20th lunar day (from June 27, 2019 to July 26, 2020) \footnote{One lunar day is approximately 28 Earth days long. The periods when LND was on are listed in \url{https://www.ieap.uni-kiel.de/et/change4/data_by_lunar_day/info}}. This was a period in the deep solar minimum between solar cycles 24 and 25. The minimum solar modulation and solar activity during that period led to the highest GCR flux since space age \citep{Fu2021ApJS}. It offers the opportunity to validate LND's performance using GCR data without the interruption by SEPs. As discussed above and in the appendix, the thresholds of several detectors had to be increased, the period we use for this work begins after all these changes have been made.

\subsection{Detection of downward-moving protons}
\label{Sec: Downward}

\subsubsection{Stopping and penetrating protons}
Stopping and penetrating protons populate DPS boxes H1 - H14, their geometry factors are shown in panel (a) of Fig.~\ref{Fig:responce-function}. The DPS boxes H1 - H9 count the protons that stop in detectors B to I. \footnote{Note that channel H9 has a relatively narrow energy range (bin width) and a smaller geometry factor than the other channels. The reason for this is an error in the definition of the corresponding DPS box (between columns 224 and 240) which does not contain the bulk of the particle population stopping in detector I. As shown in panel(a) of Fig.~\ref{Fig:2Dhistogram}, nearly half of the pixels with higher density lie outside of the black box, leading to the reduced geometry factor. This defect was not resolved in time for the flight model and hence the updated geometry factors given in the appendix should be used to make use of this channel.}
DPS boxes H10 - H14 record penetrating protons with energies above $\sim$ 35MeV. The energy bins in H10, H11, and H14 are used in this paper for the first time, they extend the measurement capability of LND beyond the energy of stopping particles. 
Unlike those for stopping particles, the geometry factors of penetrating particles partially overlap, as can be seen in panel (a) of Fig.~\ref{Fig:responce-function}. Moreover, the measurement of penetrating channels has a larger uncertainty due to uncertainties in the determination of the background as will be discussed in Sec.~\ref{sec:Modelcorrection}.
Currently, data from the penetrating channels are not yet publicly available because they still require a better calibration. With the onset of solar activity, we expect this to be achieved with some large SEP events with energies above $\sim$ 35 MeV.

\subsubsection{Calculation of Fluxes}

\begin{figure}
    \centering
    \includegraphics[width=\textwidth,]{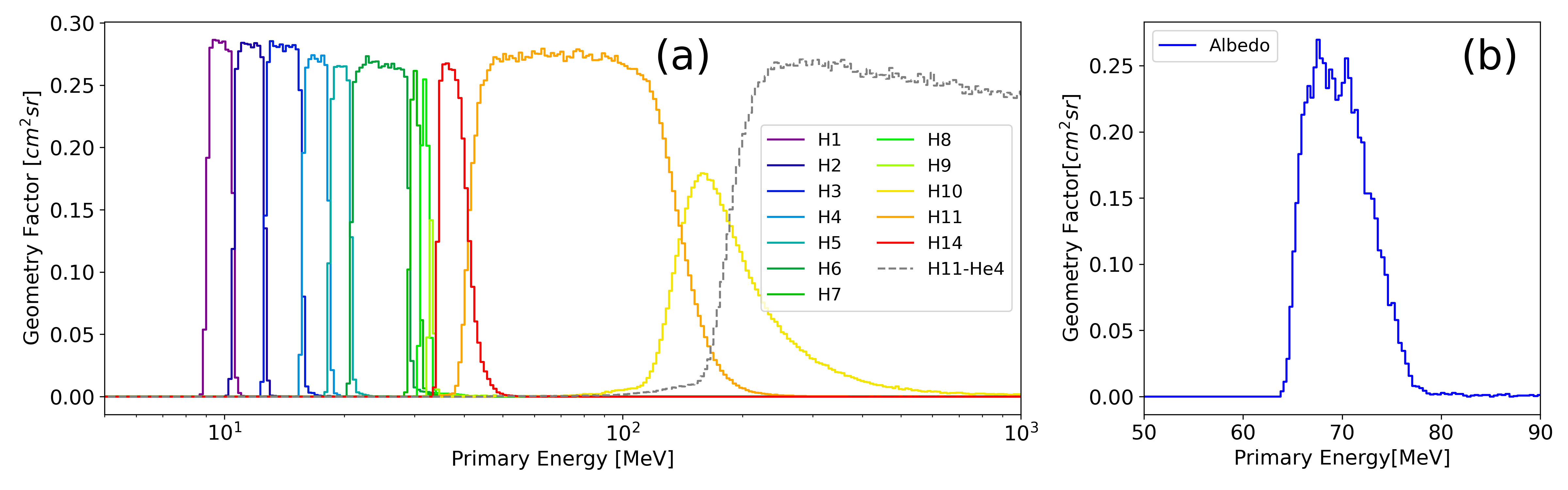}
    \caption{The geometry factors for both downward protons in (a) stopping channels (H1-H9) and penetrating ones (H10, H11, H14), and albedo protons in (b) are based on an input spectrum with power index $\gamma$=-1}.
    \label{Fig:responce-function}
\end{figure}

The differential flux, ${\rm d}F$ is calculated according to equation \ref{Eq:Flux} which is valid for all particle species, including albedo protons,
\begin{equation}
{\rm d}F = \frac{N}{{\rm d}E\cdot \bar{G} \cdot  {\rm d}t},
\label{Eq:Flux}
\end{equation}
where N is the number of counts in the DPS boxes, ${\rm d}E$ is the energy bin width determined from the response functions shown in panel (a) of Fig.~\ref{Fig:responce-function}, ${\rm d}t$ is the accumulation time of the whole measurement, and includes possible dead time corrections \citep{Wimmer2020SSR}, and $\bar{G}$ is the weighted geometry factor of each energy bin. The method that was used to calculate the averaged geometry factors is given in the appendix and their values are given in table S3.
The geometry factors of primary protons and albedo protons as a function of energy are given in panels (a) and (b) of Fig.~\ref{Fig:responce-function}.

We note that, despite the intention of the concept of geometry factors, the averaged geometry factors derived via Eq.~S1 in the supplementary material do depend on the input spectra. As explained in appendix, the contribution of high-energy particles results in non-negligible contributions in the count rates for low-energy channels. Thus, the spectral shape beyond $~\sim 100$ MeV affects the count rate in the penetrating channels. The stopping channels are less affected, as discussed in the appendix.

In table S3 we provide the averaged geometry factors derived for two different input spectra; one is a power-law spectrum with index -1 (column 3), where the particles are uniformly distributed in logarithmic energy space. The other one is the numerical CREME96 GCR model spectrum \citep{Tylka1997} for the 2019 solar minimum (column 4).

\subsubsection{Correction of the contribution from $^4$He and heavy ions to the proton flux}
\label{sec:Modelcorrection}

Close inspection of panel (a) of Fig.~\ref{Fig:2Dhistogram} shows that $^4$He and heavy ions contaminate the proton DPS boxes H1 - H9. While $^4$He is well separated from protons (its counts can be seen to populate the Xmas plot around row number 20), the Xmas plot entries between $^4$He and protons are clearly populated by a substantial background. This background which mainly affects the low-energy proton DPS boxes is primarily due to high-energy He and heavier ions which create secondary particles after they interacted with detector A. The secondary particles trigger B after which A is read out. Because the geometry factor for interaction with A is large and the flux of high-energy (GCR) ions is also high, this process happens sufficiently often to result in this background.

The penetrating channels shown in panel (b) of Fig.~\ref{Fig:2Dhistogram} are affected by the noisy detector segments A2,I1\&I2, and J1. The effect of this malfunction has been accounted for by increasing the thresholds of these detectors and disabling A2 in the LND level-3 data processing logics \cite[see][for a discussion]{Wimmer2020SSR}. This results in the asymmetry shape seen in panel (b). In the supplementary material, we give the detailed explanations of this asymmetric structure. Moreover, minimally-ionizing protons are affected more than $^4$He because of their smaller energy deposition in the detectors. Thus the detection efficiency for minimally-ionizing protons (measured in H10) is reduced because of the increased thresholds in I and J. The minimally-ionizing protons populate primarily H10, whereas minimally ionizing $^4$He populates primarily H11. Thus, as can be seen in panel (a) of Fig.~\ref{Fig:responce-function} the protons measured in H11 are strongly affected by minimally-ionizing $^4$He (dashed line in Fig.~\ref{Fig:responce-function}).

In order to quantify these different sources of contamination and reduced detection efficiencies, we ran extensive GEANT4 simulations of LND. We used the simulation setup introduced in sec.\ref{sec:simulation-set-up} to model elements from hydrogen to iron, thus including all relevant GCR species. Instead of calculating the geometry factors for detecting these elements in their instrument channels (i.e., in the appropriate boxes in the Xmas plot), we derived the geometry factors for measuring them in the proton channels (H1 - H14) as a function of energy. These simulation data were then analyzed using the current LND configuration(incl.\,thresholds). Thus we could estimate the contamination of the H1 - H14 proton channels by high-energy (GCR) helium and - to a lesser extent - heavy ions. Of course, $^4$He is rarer than hydrogen, but the contribution of GCR $^4$He to the stopping proton channels (H1 - H7) is on the order of a few percent, but larger for H8 and H9 (cf.~Tab.~S3 of supplementary material). Similarly, minimally-ionizing $^4$He affects the penetrating, but not minimally-ionizing protons in H11, only 36\% of all particles in H11 are protons (see respective entry in Tab.~S3 of supplementary material). 

To check the influence of different GCR models on the correction factors, we compare the results of the CREME 96 \citep{Tylka1997} and BON14 GCR models \citep{o2015badhwar}. The correction factors are given in columns 5 and 6 of Tab.~S3 of supplementary material. Both models have comparable values, except for the factor of channel H1, which measures protons with energy of about 10 MeV. CREME96 suggests that about 95\% of particles measured in H1 are protons, while this percentage for the BON14 is about 87\%. This discrepancy is likely due to the different spectral shapes for protons below 10 MeV. CREME96 includes low-energy (solar) particles while BON14 only includes GCR protons (and heavier ions). Thus, the ratio of "true" protons in H1 to "background protons" is larger for CREME96 than for BON14.

We note that the contamination from GCR ions discussed above is negligible during SEP events because of the higher fluxes at low energies (softness of the spectra) and the higher proton abundance compared to heavier ions. However, the same contamination processes will affect the measurements, but to a different extent that depends on the exact spectra of the solar particles. Because the GCR-induced background is always present, we can subtract the background which was measured before the SEP event from the measurements taken during the event. 

\subsection{Detection of albedo protons}
\label{Sec:upward}

\subsubsection{Measurement principle and response function}
\label{sec:albedo-response}

As alluded to in Sec.~\ref{sec:xmasplot} both downward and upward pointing particles contribute to the ``penetrating" channels in the Xmas plot between columns 246 and 266 (panel (b) of Fig.~\ref{Fig:2Dhistogram}). The extent of the $y$-axis shown covers penetrating protons and $^4$He nuclei, heavier nuclei do not contribute to the region shown here.

The \textit{x} and \textit{y} axes in panel (b) of Fig.~\ref{Fig:2Dhistogram} are defined as $ 4\cdot \mathrm{log_2(E_I/E_A)}+16$ and $4\cdot \mathrm{log_2((E_{B1}+E_C+E_D)/100)}$ respectively, where  $E_I,E_A,E_C,E_D$ represent the total energy deposited in both the inner {\em and} outer segments of the corresponding detectors, and $E_{B1}$ is the energy deposition in the inner segment of detector B.
The row number is determined by the total energy deposited by particles in detectors B1, C, and D.
The Bethe-Bloch equation \citep{bethe, bloch}, $dE \propto \frac{Z^2}{M E}$, gives the energy deposited by particles in the silicon detectors; high-energy particles tend to deposit less energy than lower-energy particles ($E$ less than $\sim$ 1 GeV) and heavy ions deposit more energy than light elements due to their higher nuclear charge, $Z$. 
Therefore, in the Xmas plot, the row number of protons with high primary energy is smaller than the low energy proton, and the protons are at the bottom of the 2D histogram. $^4$He and heavy ions are located above them.

The column number of a particle in the Xmas plot depends on the $E_I$ to $E_A$ ratio.
Suppose particles in particular the minimum ionizing particles, deposit similar energy in the front detector A and the bottom detector I, i.e. $E_I \sim E_A$, particles will be placed in the middle column of the penetrating histogram, i.e., in column 256. 
When the particle deposit more energy in detector I than detector A ($E_I > E_A$), the quantities for the x-axis are larger than 256, and those particles will be added to the pixels in the right half of panel (b). 
On the other hand, if a particle deposits less in detector I than in detector A, it will be placed in the left half of panel (b).
In principle, the former corresponds to particles moving downward, since particles first trigger detector A with larger incident energy and trigger I in the end with smaller incident energy after they lose their energy in the previous detectors, and the latter represents particles moving upward.
Here we only focus on particles with energies of up to a few tens to a few hundred MeV/nuc and for which their average energy loss in silicon detectors decreases with the increased primary energy. 

From simulations we thus determine the region marked in blue in panel (b) of upward-moving protons and calculate the corresponding geometry factor as shown in the panel (b) of Fig.~\ref{Fig:responce-function}.
The energy range of this albedo proton channel is (64.7 -- 76.7MeV), with an average geometry factor of $\mathrm{0.18\ cm^2 sr}$ for an ${\rm d} J/{\rm d}E \propto E^{-1}$ spectrum.
The energy range is defined by the 10\%-of-maximum criterion, and the uncertainty in these limits was estimated at the 5\% and 10\% of maximum and 90\% to 95\% of maximum levels. Given the steep flanks of this energy channel, these uncertainties are realistic.

Fortunately, our simulations show that the noise issue and the changes in LND configuration do not affect LND's response to albedo protons. They are too much away from the affected region in the Xmas plot, as can be seen in panel (b) of Fig.~\ref{Fig:2Dhistogram}. 

\subsubsection{Background subtraction}

\begin{figure}[!htbp]
\centering
\includegraphics[width = 0.6\textwidth]{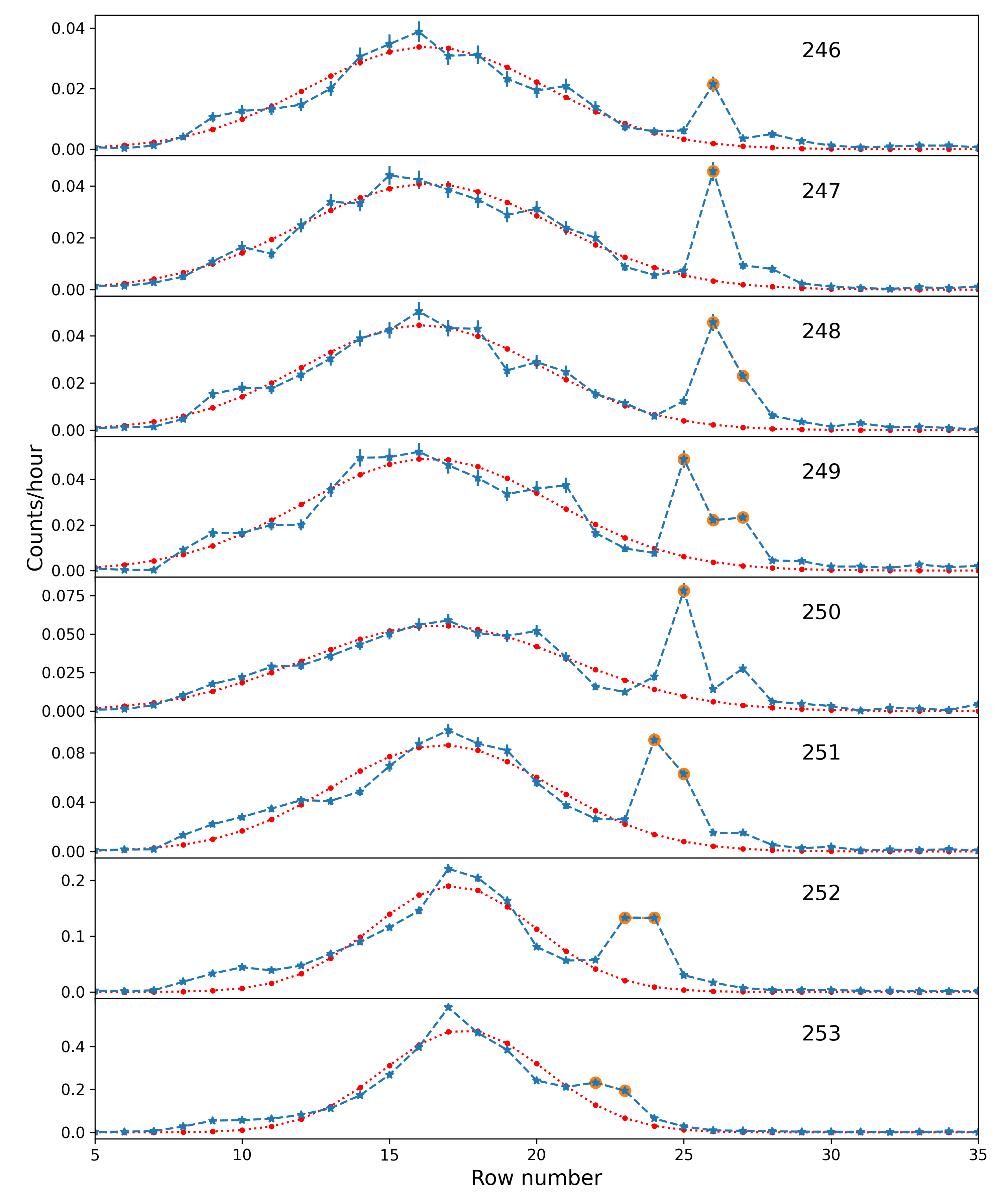}
\caption{The distribution of count rates along the Y-axis between columns 246 and 253"Xmas plot".
The blue dashed lines are the measurements and the red dashed lines are the fitted Gaussian distributions. The error bars are statistical uncertainties. The difference between the orange dots and the background gives the albedo proton-contributed count rates.}

\label{Fig:calibration_albedo}
\end{figure}

The count rates of albedo protons between 64.7 MeV and 76.7 MeV in the Xmas plots are affected by a background which is caused by high-energy protons from both directions and needs to be subtracted. 
To better visualize this background, we plot the distributions of the count rates with statistical uncertainties as error bar along the Y-axis from columns 246 to 253 in the different panels of  Fig.~\ref{Fig:calibration_albedo}. The blue dots are the averaged counts per hour in each pixel, and the orange dots are the count rates of albedo protons in the relevant pixels. 

We model the background along the y-axis with a Gaussian distribution. The fitted results are plotted as red dashed lines in Fig.~\ref{Fig:calibration_albedo}. Each panel is fitted separately with different parameters. 
Finally, the backgrounds of the albedo protons are estimated based on the fitted model and subtracted from the albedo proton data points (shown as orange data points) to determine the flux of albedo protons.

\section{Measurements and comparison with Model}
\label{sec:Measurement}

Following the calibration processes mentioned before, we derived the averaged primary proton spectrum between 8 MeV and 368 MeV and the averaged albedo proton flux between 64.7 MeV and 76.7 MeV for the lunar days between June 2019 and July 2020, i.e., when LND was on. No solar energetic protons were observed by LND  on the lunar surface.

\subsection{GCR primary proton spectrum}

The obtained LND GCR proton differential flux is plotted as empty blue circles in Fig.~\ref{Fig:spectra}. The error bars in the $x$ and $y$ directions  correspond to the widths of the energy bins and the flux uncertainty, respectively (The numerical values are also given in the last column of table S3 of appendix). The nine lowest energy channels (H1-H9) are the averages of the 1-minute data products of stopping protons in the energy range between 9 and 35 MeV. The rightmost three proton channels (from left to right, H14, H11, H10) show the penetrating proton flux with energy above 35 MeV. The detailed information of the bins is given in table S3. We only show the statistical uncertainties of the differential flux. Since each channel accumulated a large number of particles during this one-year measurement reported here, the statistical errors of each channel are small, especially for the penetrating channels. Hence the error bars of those three channels are almost invisible. Moreover, as we explained in Sec.~\ref{sec:Modelcorrection}, the contamination caused by GCR ions has been subtracted. 

Besides the measurements of LND, we also plot SOHO/EPHIN (grey square) measurements in Fig.~\ref{Fig:spectra}. SOHO/EPHIN provides measurements of protons with energies between 4.3 and 53 MeV. In this work, we use the EPHIN daily averaged level 3 proton fluxes derived from the PHA data according to the methods developed by \citet{Kuhl2020JSWSC}. Those data have 16 energy bins provided by the SOHO/EPHIN team. The periods of EPHIN data are the same as those during which LND operated, i.e., to the times when Chang'E 4 was on during local lunar daytime. One sees that the proton spectrum drops in the energy channels above 50 MeV. As explained by the EPHIN team, this change is unexpected and might not reflect the actual proton spectrum in space\footnote{P.\,K\"uhl, private communication, 2022.}.
As shown in Fig.~\ref{Fig:spectra}, the GCR proton spectrum measured by LND agrees well with that measured by SOHO/EPHIN in the energy range below 40 MeV.

Moreover, we also compare the LND measurement of GCR protons with the  CREME96 and BON14 cosmic ray models. CREME96 is a widely-used program that can simulate the ionizing-radiation environment in near-Earth space and predict the cosmic ray flux, based on the semi-empirical model of \cite{NYMMIK1992}. Their solar quiet model includes GCR, ACR, and a low-energy component below $\sim$10 MeV that originates from the sun and interplanetary space. Likewise, BON14 is based on GCR measurements from particle detectors and can predict the GCR variations related to solar modulation which is an input parameter of the model.
In Fig.~\ref{Fig:spectra}, we only plot the CREME96 cosmic ray spectra of 2019 as dashed blue lines since the BON14 predicts a very similar proton flux during solar minimum. The lower curve is the original spectrum generated by the CREME website\href{https://creme.isde.vanderbilt.edu/CREME-MC} for this period. Obviously, the GCR model underestimates the proton flux at a few tens of MeV during solar minimum 24/25, and the measurement is about 48\%$\pm$2\% higher than the model spectra. This percentage is the averaged ratio between the measurement and the mode. The comparison with BON14 leads to the same conclusion. Therefore, we multiplied the CREME spectrum by a factor of 1.56 for the limited energy range which fits the LND measurement best. This implies that the actual solar modulation during the solar minimum 24/25 may be weaker than the modulation used in numerical models. 
A similar disagreement between models and experimental data has been reported by \cite{Mrigakshi2012JGR} during the solar minimum 23/24 and further discussed by \cite{Matthia2013}. They discuss that the averaged sunspot number is not an accurate predictor of GCR fluxes during solar minimum and show substantial differences between the GCR ISO model \citep{iso200415390} (which is a later version of the \cite{NYMMIK1992} model) and the BON model. In fact, solar modulation has been reported to be reduced in 2019-2020 and GCR intensities at 1 au during this period reached the highest record in the space age \citep{Fu2021ApJS}.

\subsection{Albedo protons}

\begin{figure}[!htbp]
\centering
\includegraphics[width = 0.9\textwidth]{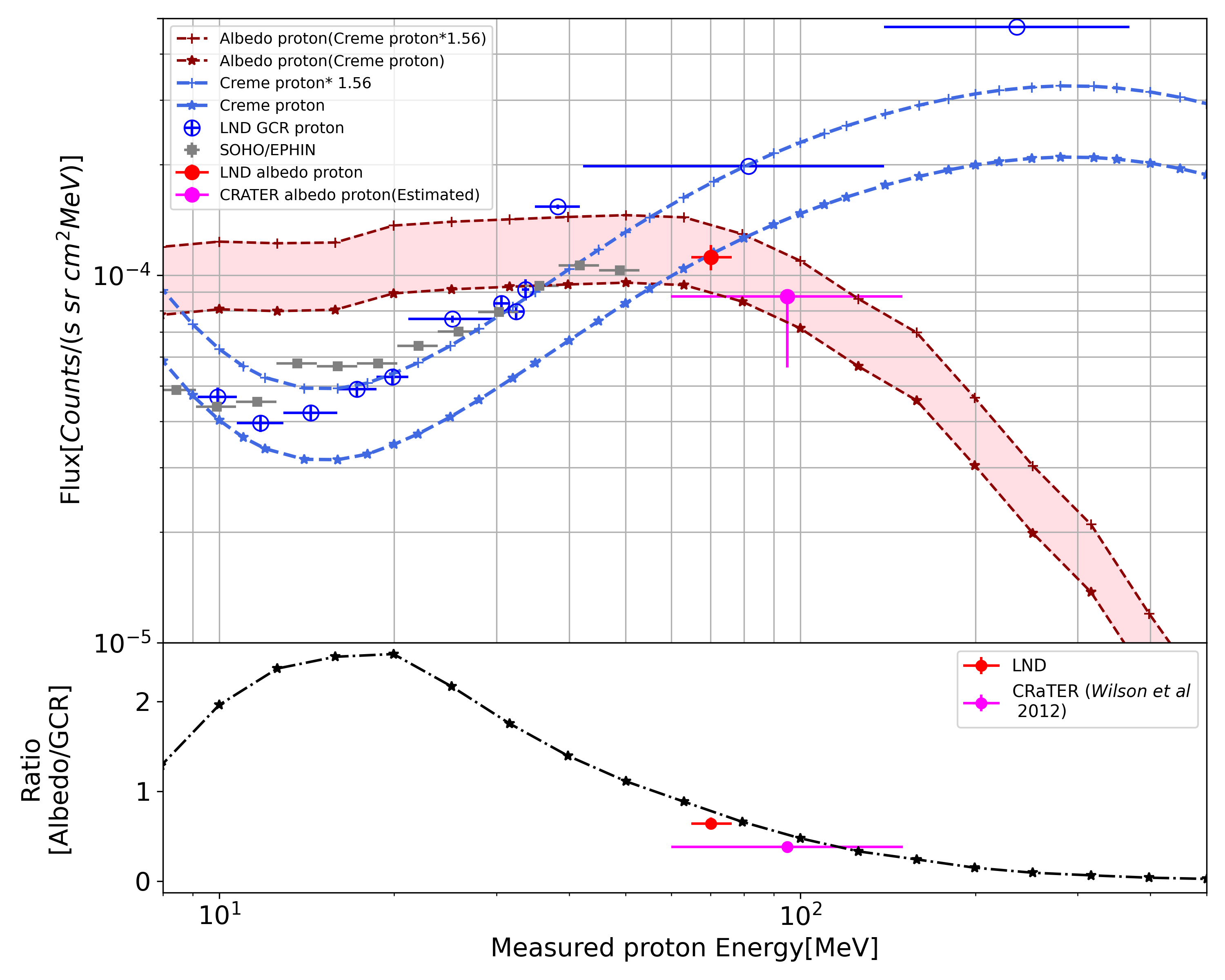}
\caption{(Top) The GCR primary proton spectra including both measurements (LND in empty blue circles and SOHO/EPHIN in grey triangles) and modeled results (blue dashed lines). Albedo proton spectra are drawn in reddish colors. The simulation results from REDMoon are within the transparent red area. The LND measurement of albedo protons is the red circle. The error bar is the systematic uncertainty as defined in sec.~{\ref{Sec:upward}}. The estimated albedo proton flux that we derived using the ratio of albedo to primary proton measured by LRO/CRaTER is the magenta data point. (Bottom) The ratio of albedo to primary flux versus proton energy. The red and magenta points are the ratio of albedo to primary proton flux derived from LND data in this study and that from LRO/CRaTER data \citep{Wilson2012}, respectively. The black line shows the prediction from the REDMoon model. The detailed explanations are given in the main text, the values for the LND data pints are given in Tab.~S3 of supplementary material.}

\label{Fig:spectra}
\end{figure}

As discussed in Sec.~\ref{Sec:upward}, LND can resolve albedo protons and determine their flux between  64.7 and 76.7 MeV. This is shown as the red circle in Fig.~\ref{Fig:spectra} for the same time period as the GCR spectra.
The background has been subtracted and the error of the flux is the resulting systematic uncertainty which is larger than the statistical error. The averaged flux at $\sim$ 70.5 MeV is about $1.12 \mathrm{\pm 0.09\times 10^{-4} (cm^2\ s\ sr\ MeV)^{-1}}$  which is remarkably close to the primary GCR proton flux measured by LND in this energy range. Albedo protons are the product of high-energy GCRs which interact with the lunar soil. A small fraction of the reaction products can escape from the soil and thus forms the albedo population. This interaction is modeled by the REDMoon code \citep{DobyndeREDMOON2021} which then predicts the lunar proton albedo as shown by the red lines in Fig.~\ref{Fig:spectra} and discussed in the following paragraph.

\begin{figure}[!htbp]
\centering
\includegraphics[width = 0.95\textwidth]{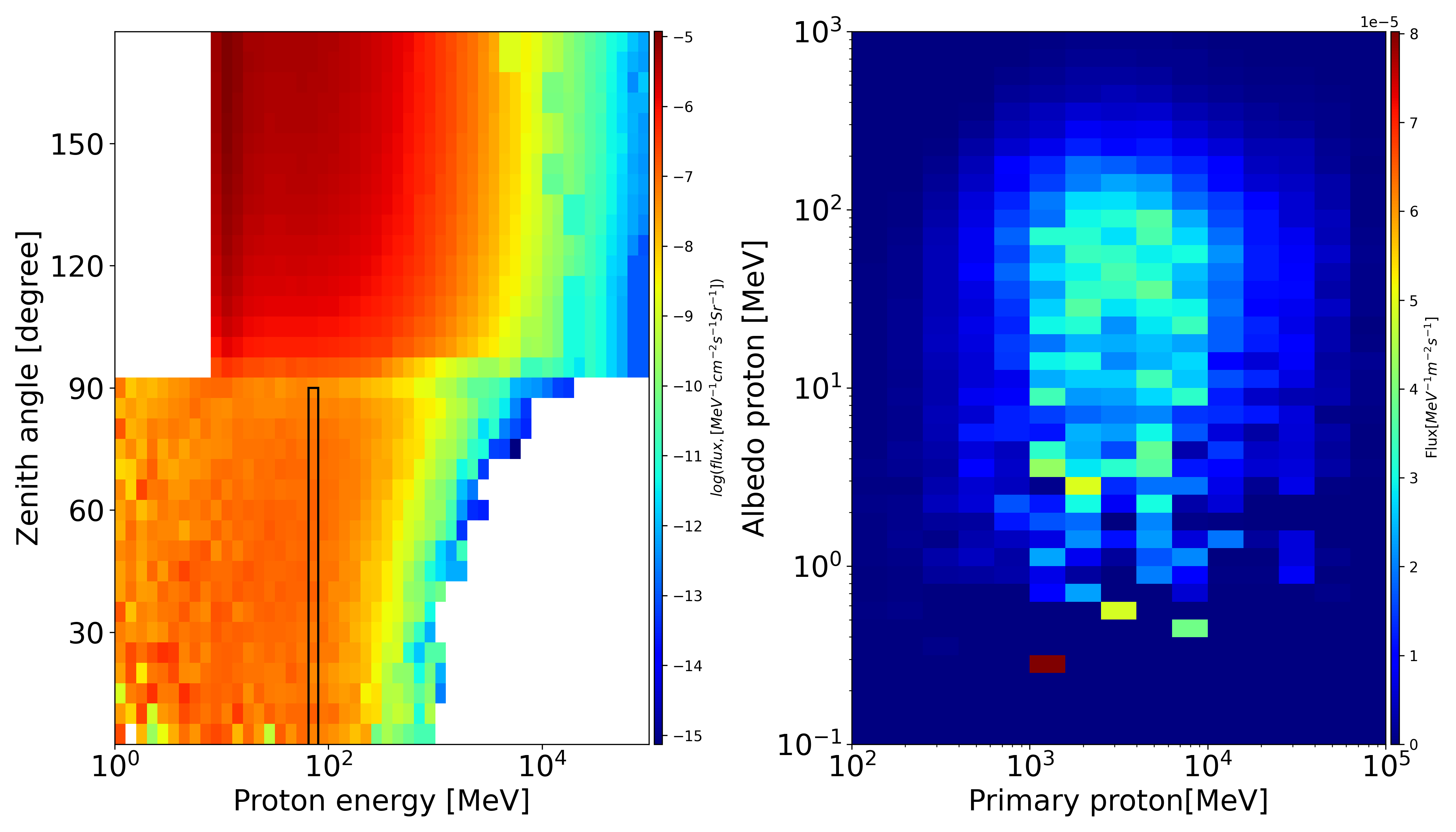}
\caption{(left) The angular distribution of protons on the lunar surface including both albedo protons (0-90 degree) and GCR primary protons (90 - 180 degree). The plot is modified from panel (a) of Fig. 4 in \citet{DobyndeREDMOON2021}. (right): Albedo proton flux as a function of primary protons shows a maximum around albedo energies of 80 MeV. These albedo particles are produced by primary protons with much higher energies. Both albedo particle distributions are derived by REDMoon model.}
\label{Fig:distribution}
\end{figure}

Here the REDMoon model takes the CREME GCR model as input to calculate the angle- and energy-resolved surface particle flux on the Moon. The result for the time period reported here is given in Fig.~\ref{Fig:distribution} which shows the isotropically distributed downward flux (zenith angle 90-180 deg) at the lunar surface in the upper half of the figure. The lower half shows the angle- and energy-resolved upward flux of (albedo) protons, the black rectangle marks the energy range which LND measures. The albedo protons are primarily due to GCR protons, but GCR helium and heavier ions contribute a small fraction. The REDMoon results shown as dashed red lines in Fig.~\ref{Fig:spectra} were calculated with the unscaled CREME model as well as a CREME input spectrum scaled by a constant factor of 1.56, as discussed above. The region between these two model predictions is shaded in pink and represents the upward proton flux that is expected to be measured by LND. 
The decrease of the albedo protons flux at energy higher than $\sim$100~MeV reflects that the generation efficiency of secondary particles drops with energy.

The lower panel of Fig.~\ref{Fig:spectra} shows the ratio of upward (albedo) to downward (GCR) proton flux as derived by taking the ratio of REDMoon and CREME simulation results. Note that the ratio is the same for the scaled and non-scaled inputs of CREME spectra and that it is computed at the same energy of downward and upward propagating protons. The ratio peaks around 20 MeV, where the albedo protons are more than two times higher than the primary GCR protons in that energy range. Above it, the ratio drops as the GCR flux increases and the albedo flux decreases. The high ratio ($>1$) is the consequence of the high-energy GCR making the main contribution to the low-energy albedo protons.

We determined the ratio of albedo protons to the primary protons in the energy range between 64.7 and 76.7 MeV, as measured by LND. The average primary proton flux is estimated from the scaled GCR spectrum that fits the LND measurements. We found that the ratio is about 0.64$\pm$0.07 which means that the albedo protons are a significant contribution to the particle flux on the lunar surface in this energy range. Thus, about 39\% $\pm$6\% of all protons around 70 MeV are secondary albedo particles emitted from the lunar regolith. 
The second data point shown in magenta dots is the ratio reported by the CRaTER team \citep{Wilson2012}. The REDMoon simulation results agree well with the both the LND and CRATER measurements in their energy ranges.

The CRaTER data point was only reported by \citet{Wilson2012} as a ratio of the number of upward (albedo) protons to that of downward protons measured within the same FOV and the same energy range for both particle populations. They found a value of 0.38$\pm$0.02 for this ratio which is different from the one reported by LND in this work. It is important to note that the CRaTER measurements are sensitive to a different energy range than LND, i.e., between 60 and 150 MeV (with an average energy higher than the LND-seen albedo protons) and were obtained between 2009 to 2011. This time period also corresponds to a solar minimum and we don't expect the ratio of upward to downward protons to depend strongly on the detailed differences in the two solar minima between solar cycles 23/24 and 24/25. 

Primary (GCR) protons interact with the lunar surface and can produce the albedo population reported here. \cite{Schwadron2017} discuss the various generation mechanisms in more detail. The overall shape of the ratio of albedo to primary proton can be understood without sophisticated modeling: Albedo protons necessarily must have lost a significant part of their energy in the uppermost layers of the lunar regolith. Because the primary (GCR) proton flux increases with energy in and beyond the energy range covered by LND, the ratio of albedo/primary proton flux must decrease with energy. This simple explanation is supported by REDMoon model calculations and were also obtained by \cite{Looper2013SpWea}
in their simulation of the radiation environment on the lunar orbiter of 50km height. This also explains why the LND observations of the albedo to primary ratio is nearly two times higher than that of LRO/CRaTER.
In Tab.~\ref{Tab:ratio}, we list the ratios of different cases, including the ratios from LND in 2019, CRaTER in 2009, and the ratio from the simulation averaged over different energy. We always use the GCR spectrum of 2019 as the input spectrum in the simulation. Ignoring the time difference, we can directly compare the simulation with measurements in the same energy range. The modeled albedo to primary proton ratio averaged between in the LND energy range (65-76 MeV) is 0.8, which is slightly higher than LND's measurement, but the albedo flux falls within the pink shaded band in the upper panel. For CRaTER one needs to average the model over a wider energy range which is given by CRaTER's energy range. Thus folding the model function with the CRaTER energy range, we obtain a value of 0.33, which is comparable with the CRaTER's result.

\cite{Wilson2012} only report the albedo to primary proton ratio, but not the upward proton flux. This can easily be estimated by multiplying the scaled CREME model flux in the appropriate energy range with the reported albedo ratio and is shown as the magenta point in the upper panel of Fig.~\ref{Fig:spectra}. The albedo proton flux that is calculated using the original (unscaled) GCR spectrum can be taken as a lower limit. LND measurements, CRaTER measurements, and the REDMoon simulation are consistent within the reported uncertainties.

\begin{table}[!htbp]
\centering
\caption{The ratio of lunar albedo protons to primary protons.}
\begin{tabular}{c c c c}        
\hline\hline                 
LND(64.7-76.7MeV) &  CRaTER(60-150MeV) &  simulation(65-76MeV) & simulation(60-150MeV)\\    
\hline
2019-2020 & 2008-2009& 2019 & 2019  \\
\hline                        
64$\%\pm 7$  &   38$\%\pm$2   & 80$\%$  & 33$\%$ \\
\hline                                   
\end{tabular}
\label{Tab:ratio}
\end{table}

\section{Summary, Discussion and conclusion}
\label{sec:conclusions}

We have performed and reported a careful re-analysis of the LND calibration. LND data in the energy range between 9 MeV and 368 MeV agrees well with data from the SOHO/EPHIN instrument. LND data up to 35 MeV are for well-resolved protons stopping in LND's detector stack; the remaining three data points are for penetrating protons.

Furthermore, we compared LND's proton spectrum with the CREME96 GCR model. We find that the CREME96 predictions underestimate the proton flux during reported time period in the solar minimum 24/25. The average ratio between LND measurements and the CREME96 model predictions is $\sim$1.48$\pm$0.02. This may be due to the much weaker solar modulation during the deep solar minimum 24/25. In fact, GCR intensities at 1 au during this time period reached the highest recorded fluxes in the space age \citep{Fu2021ApJS}.

With its stack of 10 silicon detectors, LND is capable of measuring both the downward as well as the upward proton flux. Such albedo protons are the result of the interaction of high-energy GCR particles with the lunar regolith. The bulk of the albedo particles is created by the high-energy portion of the primary GCR protons, as shown in the right hand panel of Fig.~\ref{Fig:distribution}. This part of the GCR spectrum is less affected by solar modulation than the low-energy part. Therefore, when considering the effect of solar modulation on the ratio of upward to downward particles one needs to consider that this ratio is calculated at the same primary energy of the upward and downward-pointing particles. Because the low-energy part of the GCR spectrum is strongly affected by solar modulation, the up/down ratio is primarily affected by the low-energy downward flux and much less so by the upward flux which is primarily determined by the high-energy downward-pointing flux, as reported by \cite{Wilson2012}. This effect is especially important during solar particle events. 
Comparing the measured upward to downward  differential fluxes we find a ratio of $0.64 \pm 0.07$ in the energy range between $64.7 - 76.7$ MeV. This ratio is shown as the red data point in the lower panel of Fig.~\ref{Fig:spectra}. Because of the additional shielding from beneath provided by the Chang'E 4 lander, the energy ranges resolved by LND differ for the upward and downward particle populations. 
Furthermore, we scaled the upward to downward flux ratio reported by the CRaTER team using data prior to solar minimum 23/24 to the time period investigated here. We used the scaled CREME96 downward flux to determine the upward flux expected in the CRaTER energy range. The resulting upward flux (shown as the magenta data point in the lower panel of Fig.~\ref{Fig:spectra}) agrees well with LND's measurement and the REDMoon predictions, despite being valid for a different energy range ($60 - 150$ MeV).

As already stated, the REDMoon simulation tool provides the upward (albedo) differential proton flux. We compare the REDMoon model results with the downward flux from the scaled CREME96 model to derive the ratio of upward to downward differential flux shows as the dash-dotted line in the lower panel of Fig.~\ref{Fig:spectra}. At energies below $\sim$ 50 MeV, the ratio is above 1 and it has a peak value above 2 near 20 MeV. With increasing energy, the ratio decreases as the generation of albedo protons at this energy decreases. We observe that the modeled energy dependence of the albedo proton flux is confirmed by the combination of the LND and (scaled) CRaTER data.


LND is the first high-energy charged-particles telescope to be working on the lunar surface. We have presented measurements of low-energy cosmic ray protons in conjunction with albedo protons and that the latter contribute significantly to the particle flux on the lunar surface. 
We also show that the albedo proton flux is energy dependent, and that both primary and albedo proton fluxes are in good agreement with measurements from LRO/CRaTER. 
Obviously, due to their high flux and relatively low energy, albedo protons are an important contribution to the radiation to which astronauts would be exposed to on the surface of the Moon. Albedo protons are energetic enough to penetrate a space suit and have sufficiently low energy do stop in the body of an astronaut, thus depositing all their energy in the astronaut's tissue. 
The Sun is now becoming more active and future data from large solar particle events will be critical to understand their effect on the lunar radiation environment and its effect on human exploration of the Moon.

\section*{Conflict of Interest Statement}

The authors declare that the research was conducted in the absence of any commercial or financial relationships that could be construed as a potential conflict of interest.

\section*{Author Contributions}

ZX led the data analysis and paper writing. RW and JG advised on the data analysis and improved the paper writing. MD provided the simulation data of the REDMoon model, and PK offered SOHO/EPHIN data. SK and SZ provided valuable information and suggestions. All authors read the paper and approved its submission.

\section*{Funding}
The Lunar Lander Neutron and Dosimetry (LND) instrument was funded by the German Space Agency, DLR, and its space 
Administration under grant 50 JR 1604 to the Christian-Albrechts-University (CAU) Kiel as well 
as the Beijing Municipal Science and Technology Commission (grant no. Z181100002918003) 
and NSFC (grant no. 41941001). JG and MD are supported by the Strategic Priority Program of the Chinese Academy of Sciences (Grant No. XDB41000000) and the National Natural Science Foundation of China (Grant No. 42074222, 41941001). 
We acknowledge financial support by Land Schleswig-Holstein within the funding programme Open Access Publikationsfonds.

\section*{Acknowledgments}
We thank the two anonymous reviewers for their thorough reviews which helped to improve this paper and the many individuals who were involved in conceiving, designing, building, testing, and launching LND and Chang'E 4.

\section*{Supplemental Data}
The data calibrations and the LND configuration changes are given in the supplementary material.

\section*{Data Availability Statement}
The scientific data of Lunar Lander Neutron and Dosimetry Experiment are provided by China National Space Administration.


\bibliographystyle{Frontiers-Harvard} 
\bibliography{main}

\begin{thebibliography}{32}
\providecommand{\natexlab}[1]{#1}
\expandafter\ifx\csname urlstyle\endcsname\relax
  \providecommand{\doi}[1]{doi:\discretionary{}{}{}#1}\else
  \providecommand{\doi}{doi:\discretionary{}{}{}\begingroup
  \urlstyle{rm}\Url}\fi
\providecommand{\selectlanguage}[1]{\relax}
\providecommand{\bibAnnoteFile}[1]{%
  \IfFileExists{#1}{\begin{quotation}\noindent\textsc{Key:} #1\\
  \textsc{Annotation:}\ \input{#1}\end{quotation}}{}}
\providecommand{\bibAnnote}[2]{%
  \begin{quotation}\noindent\textsc{Key:} #1\\
  \textsc{Annotation:}\ #2\end{quotation}}

\bibitem[{Agostinelli et~al.(2003)Agostinelli, Allison, Amako, Apostolakis,
  Araujo, Arce et~al.}]{Agostinelli2003}
Agostinelli, S., Allison, J., Amako, K., Apostolakis, J., Araujo, H., Arce, P.,
  et~al. (2003).
\newblock {GEANT4 - A simulation toolkit}.
\newblock \emph{Nuclear Instruments and Methods in Physics Research, Section A:
  Accelerators, Spectrometers, Detectors and Associated Equipment} 506,
  250--303.
\newblock \doi{10.1016/S0168-9002(03)01368-8}
\bibAnnoteFile{Agostinelli2003}

\bibitem[{{Appel} et~al.(2018){Appel}, {K{\"o}ehler}, {Guo}, {Ehresmann},
  {Zeitlin}, {Matthi{\"a}} et~al.}]{2018Appel}
{Appel}, J.~K., {K{\"o}ehler}, J., {Guo}, J., {Ehresmann}, B., {Zeitlin}, C.,
  {Matthi{\"a}}, D., et~al. (2018).
\newblock {Detecting Upward Directed Charged Particle Fluxes in the Mars
  Science Laboratory Radiation Assessment Detector}.
\newblock \emph{Earth and Space Science} 5, 2--18.
\newblock \doi{10.1002/2016EA000240}
\bibAnnoteFile{2018Appel}

\bibitem[{{Berger} et~al.(2020){Berger}, {Matthi{\"a}}, {Burmeister},
  {Zeitlin}, {Rios}, {Stoffle} et~al.}]{Berger2020JSWSC}
{Berger}, T., {Matthi{\"a}}, D., {Burmeister}, S., {Zeitlin}, C., {Rios}, R.,
  {Stoffle}, N., et~al. (2020).
\newblock {Long term variations of galactic cosmic radiation on board the
  International Space Station, on the Moon and on the surface of Mars}.
\newblock \emph{Journal of Space Weather and Space Climate} 10, 34.
\newblock \doi{10.1051/swsc/2020028}
\bibAnnoteFile{Berger2020JSWSC}

\bibitem[{Bethe(1930)}]{bethe}
Bethe, H.~A. (1930).
\newblock Zur {T}heorie des {D}urchgangs schneller {K}orpuskularstrahlen durch
  {M}aterie.
\newblock \emph{Ann. d. Phys.} 5, 325
\bibAnnoteFile{bethe}

\bibitem[{Bloch(1933)}]{bloch}
Bloch, F. (1933).
\newblock Zur {B}remsung rasch bewegter {T}eilchen beim {D}urchgang durch
  {M}aterie.
\newblock \emph{Ann. d. Phys} 16, 285
\bibAnnoteFile{bloch}

\bibitem[{{Cucinotta} and {Chappell}(2011)}]{cucinotta-etal-2011}
{Cucinotta}, F.~A. and {Chappell}, L.~J. (2011).
\newblock {Updates to Astronaut Radiation Limits: Radiation Risks for
  Never-Smokers}.
\newblock \emph{Radiation Research} 176, 102--114.
\newblock \doi{10.1667/RR2540.1}
\bibAnnoteFile{cucinotta-etal-2011}

\bibitem[{{Dobynde} and {Guo}(2021)}]{DobyndeREDMOON2021}
{Dobynde}, M.~I. and {Guo}, J. (2021).
\newblock {Radiation Environment at the Surface and Subsurface of the Moon:
  Model Development and Validation}.
\newblock \emph{Journal of Geophysical Research (Planets)} 126, e06930.
\newblock \doi{10.1029/2021JE006930}
\bibAnnoteFile{DobyndeREDMOON2021}

\bibitem[{{Dorman}(2004)}]{Dorman2004ASSL}
{Dorman}, L.~I. (2004).
\newblock \emph{{Cosmic Rays in the Earth's Atmosphere and Underground}}, vol.
  303.
\newblock \doi{10.1007/978-1-4020-2113-8}
\bibAnnoteFile{Dorman2004ASSL}

\bibitem[{{Fu} et~al.(2021){Fu}, {Zhang}, {Zhao}, and {Li}}]{Fu2021ApJS}
{Fu}, S., {Zhang}, X., {Zhao}, L., and {Li}, Y. (2021).
\newblock {Variations of the Galactic Cosmic Rays in the Recent Solar Cycles}.
\newblock \emph{apjs} 254, 37.
\newblock \doi{10.3847/1538-4365/abf936}
\bibAnnoteFile{Fu2021ApJS}

\bibitem[{{Guo} et~al.(2019){Guo}, {Banjac}, {R{\"o}stel}, {Terasa}, {Herbst},
  {Heber} et~al.}]{Guo2019JSWSC}
{Guo}, J., {Banjac}, S., {R{\"o}stel}, L., {Terasa}, J.~C., {Herbst}, K.,
  {Heber}, B., et~al. (2019).
\newblock {Implementation and validation of the GEANT4/AtRIS code to model the
  radiation environment at Mars}.
\newblock \emph{Journal of Space Weather and Space Climate} 9, A2.
\newblock \doi{10.1051/swsc/2018051}
\bibAnnoteFile{Guo2019JSWSC}

\bibitem[{{Hassler} et~al.(2012){Hassler}, {Zeitlin}, {Wimmer-Schweingruber},
  {B{\"o}ttcher}, {Martin}, {Andrews} et~al.}]{Hassler2012SSRv}
{Hassler}, D.~M., {Zeitlin}, C., {Wimmer-Schweingruber}, R.~F., {B{\"o}ttcher},
  S., {Martin}, C., {Andrews}, J., et~al. (2012).
\newblock {The Radiation Assessment Detector (RAD) Investigation}.
\newblock \emph{ssr} 170, 503--558.
\newblock \doi{10.1007/s11214-012-9913-1}
\bibAnnoteFile{Hassler2012SSRv}

\bibitem[{ISO(2004)}]{iso200415390}
ISO, I. (2004).
\newblock 15390: 2004: Space environment (natural and artifical)-galactic
  cosmic ray model.
\newblock \emph{Geneva, Switzerland: International Organization for
  Standardization}
\bibAnnoteFile{iso200415390}

\bibitem[{{K{\"u}hl} et~al.(2020){K{\"u}hl}, {Heber}, {G{\'o}mez-Herrero},
  {Malandraki}, {Posner}, and {Sierks}}]{Kuhl2020JSWSC}
{K{\"u}hl}, P., {Heber}, B., {G{\'o}mez-Herrero}, R., {Malandraki}, O.,
  {Posner}, A., and {Sierks}, H. (2020).
\newblock {The Electron Proton Helium INstrument as an example for a Space
  Weather Radiation Instrument}.
\newblock \emph{Journal of Space Weather and Space Climate} 10, 53.
\newblock \doi{10.1051/swsc/2020056}
\bibAnnoteFile{Kuhl2020JSWSC}

\bibitem[{Looper et~al.(2013)Looper, Mazur, Blake, Spence, Schwadron, Golightly
  et~al.}]{Looper2013}
Looper, M.~D., Mazur, J.~E., Blake, J.~B., Spence, H.~E., Schwadron, N.~A.,
  Golightly, M.~J., et~al. (2013).
\newblock The radiation environment near the lunar surface: Crater observations
  and geant4 simulations.
\newblock \emph{Space Weather} 11, 142--152.
\newblock \doi{10.1002/swe.20034}
\bibAnnoteFile{Looper2013}

\bibitem[{{Looper} et~al.(2013){Looper}, {Mazur}, {Blake}, {Spence},
  {Schwadron}, {Golightly} et~al.}]{Looper2013SpWea}
{Looper}, M.~D., {Mazur}, J.~E., {Blake}, J.~B., {Spence}, H.~E., {Schwadron},
  N.~A., {Golightly}, M.~J., et~al. (2013).
\newblock {The radiation environment near the lunar surface: CRaTER
  observations and Geant4 simulations}.
\newblock \emph{Space Weather} 11, 142--152.
\newblock \doi{10.1002/swe.20034}
\bibAnnoteFile{Looper2013SpWea}

\bibitem[{{Matthi{\"a}} et~al.(2013){Matthi{\"a}}, {Berger}, {Mrigakshi}, and
  {Reitz}}]{Matthia2013}
{Matthi{\"a}}, D., {Berger}, T., {Mrigakshi}, A.~I., and {Reitz}, G. (2013).
\newblock {A ready-to-use galactic cosmic ray model}.
\newblock \emph{Advances in Space Research} 51, 329--338.
\newblock \doi{10.1016/j.asr.2012.09.022}
\bibAnnoteFile{Matthia2013}

\bibitem[{{McKinney} et~al.(2006){McKinney}, {Lawrence}, {Prettyman}, {Elphic},
  {Feldman}, and {Hagerty}}]{mckinney-etal-2006}
{McKinney}, G.~W., {Lawrence}, D.~J., {Prettyman}, T.~H., {Elphic}, R.~C.,
  {Feldman}, W.~C., and {Hagerty}, J.~J. (2006).
\newblock {MCNPX benchmark for cosmic ray interactions with the Moon}.
\newblock \emph{Journal of Geophysical Research (Planets)} 111, E06004.
\newblock \doi{10.1029/2005JE002551}
\bibAnnoteFile{mckinney-etal-2006}

\bibitem[{Mitrofanov et~al.(2016)Mitrofanov, Sanin, and
  Litvak}]{Mitrofanov2016}
Mitrofanov, I.~G., Sanin, A.~B., and Litvak, M.~L. (2016).
\newblock {Water in the Moon's polar areas: Results of LEND neutron telescope
  mapping}.
\newblock \emph{Doklady Physics} 61, 98--101.
\newblock \doi{10.1134/S1028335816020117}
\bibAnnoteFile{Mitrofanov2016}

\bibitem[{{Mrigakshi} et~al.(2012){Mrigakshi}, {Matthi{\"a}}, {Berger},
  {Reitz}, and {Wimmer-Schweingruber}}]{Mrigakshi2012JGR}
{Mrigakshi}, A.~I., {Matthi{\"a}}, D., {Berger}, T., {Reitz}, G., and
  {Wimmer-Schweingruber}, R.~F. (2012).
\newblock {Assessment of galactic cosmic ray models}.
\newblock \emph{Journal of Geophysical Research (Space Physics)} 117, A08109.
\newblock \doi{10.1029/2012JA017611}
\bibAnnoteFile{Mrigakshi2012JGR}

\bibitem[{M{\"u}ller-Mellin et~al.(1995)M{\"u}ller-Mellin, Kunow, Flei{\ss}ner,
  Pehlke, Rode, R{\"o}schmann et~al.}]{muller1995costep}
M{\"u}ller-Mellin, R., Kunow, H., Flei{\ss}ner, V., Pehlke, E., Rode, E.,
  R{\"o}schmann, N., et~al. (1995).
\newblock Costep-comprehensive suprathermal and energetic particle analyser.
\newblock \emph{Solar Physics} 162, 483--504
\bibAnnoteFile{muller1995costep}

\bibitem[{Nymmik et~al.(1992)Nymmik, Panasyuk, Pervaja, and
  Suslov}]{NYMMIK1992}
Nymmik, R., Panasyuk, M., Pervaja, T., and Suslov, A. (1992).
\newblock A model of galactic cosmic ray fluxes.
\newblock \emph{International Journal of Radiation Applications and
  Instrumentation. Part D. Nuclear Tracks and Radiation Measurements} 20,
  427--429.
\newblock \doi{https://doi.org/10.1016/1359-0189(92)90028-T}.
\newblock Special Section Galactic Cosmic Radiation: Constraints on Space
  Exploration
\bibAnnoteFile{NYMMIK1992}

\bibitem[{O'Neill et~al.(2015)O'Neill, Golge, and Slaba}]{o2015badhwar}
O'Neill, P., Golge, S., and Slaba, T. (2015).
\newblock \emph{Badhwar-O'Neill 2014 galactic cosmic ray flux model
  description}.
\newblock Tech. rep.
\bibAnnoteFile{o2015badhwar}

\bibitem[{Schwadron et~al.(2017)Schwadron, Wilson, Jordan, Looper, Zeitlin,
  Townsend et~al.}]{Schwadron2017}
Schwadron, N.~A., Wilson, J.~K., Jordan, A.~P., Looper, M.~D., Zeitlin, C.,
  Townsend, L.~W., et~al. (2017).
\newblock Using proton radiation from the moon to search for diurnal variation
  of regolith hydrogenation.
\newblock \emph{Planetary and Space Science} 162, 113--132.
\newblock \doi{10.1016/j.pss.2017.09.012}
\bibAnnoteFile{Schwadron2017}

\bibitem[{Schwadron et~al.(2016)Schwadron, Wilson, Looper, Jordan, Spence,
  Blake et~al.}]{Schwadron2016}
Schwadron, N.~A., Wilson, J.~K., Looper, M.~D., Jordan, A.~P., Spence, H.~E.,
  Blake, J.~B., et~al. (2016).
\newblock Signatures of volatiles in the lunar proton albedo.
\newblock \emph{Icarus} \doi{10.1016/j.icarus.2015.12.003}
\bibAnnoteFile{Schwadron2016}

\bibitem[{Spence et~al.(2010)Spence, Case, Golightly, Heine, Larsen, Blake
  et~al.}]{spence2010}
Spence, H.~E., Case, A.~W., Golightly, M.~J., Heine, T., Larsen, B.~A., Blake,
  J.~B., et~al. (2010).
\newblock Crater: The cosmic ray telescope for the effects of radiation
  experiment on the lunar reconnaissance orbiter mission.
\newblock \emph{Space Science Reviews} 150, 243--284.
\newblock \doi{10.1007/s11214-009-9584-8}
\bibAnnoteFile{spence2010}

\bibitem[{Spence et~al.(2013)Spence, Golightly, Joyce, Looper, Schwadron, Smith
  et~al.}]{Spence2013}
Spence, H.~E., Golightly, M.~J., Joyce, C.~J., Looper, M.~D., Schwadron, N.~A.,
  Smith, S.~S., et~al. (2013).
\newblock Relative contributions of galactic cosmic rays and lunar proton
  "albedo" to dose and dose rates near the moon.
\newblock \emph{Space Weather} 11, 643--650.
\newblock \doi{10.1002/2013SW000995}
\bibAnnoteFile{Spence2013}

\bibitem[{{Treiman}(1953)}]{TREIMAN1953PhRv}
{Treiman}, S.~B. (1953).
\newblock {The Cosmic-Ray Albedo}.
\newblock \emph{Physical Review} 91, 957--959.
\newblock \doi{10.1103/PhysRev.91.957}
\bibAnnoteFile{TREIMAN1953PhRv}

\bibitem[{Tylka et~al.(1997)Tylka, Adams, Boberg, Brownstein, Dietrich,
  Flueckiger et~al.}]{Tylka1997}
Tylka, A.~J., Adams, J.~H., Boberg, R., Brownstein, B., Dietrich, W.~F.,
  Flueckiger, E.~O., et~al. (1997).
\newblock Creme96: A revision of the cosmic ray e [ects on micro-electronics
  code.
\newblock \emph{IEEE TRANSACTIONS ON NUCLEAR SCIENCE} 44
\bibAnnoteFile{Tylka1997}

\bibitem[{Wilson et~al.(2012)Wilson, Spence, Kasper, Golightly, Blake, Mazur
  et~al.}]{Wilson2012}
Wilson, J.~K., Spence, H.~E., Kasper, J., Golightly, M., Blake, J.~B., Mazur,
  J.~E., et~al. (2012).
\newblock The first cosmic ray albedo proton map of the moon.
\newblock \emph{Journal of Geophysical Research E: Planets} 117, 1--7.
\newblock \doi{10.1029/2011JE003921}
\bibAnnoteFile{Wilson2012}

\bibitem[{Wimmer-Schweingruber et~al.(2020)Wimmer-Schweingruber, Yu,
  B{\"{o}}ttcher, Zhang, Burmeister, Lohf et~al.}]{Wimmer2020SSR}
Wimmer-Schweingruber, R.~F., Yu, J., B{\"{o}}ttcher, S.~I., Zhang, S.,
  Burmeister, S., Lohf, H., et~al. (2020).
\newblock {The Lunar Lander Neutron and Dosimetry (LND) Experiment on Chang'E
  4}.
\newblock \emph{Space Science Reviews} 216, 104.
\newblock \doi{10.1007/s11214-020-00725-3}
\bibAnnoteFile{Wimmer2020SSR}

\bibitem[{{Xu} et~al.(2020){Xu}, {Guo}, {Wimmer-Schweingruber}, {Freiherr von
  Forstner}, {Wang}, {Dresing} et~al.}]{Xu2020ApJ}
{Xu}, Z., {Guo}, J., {Wimmer-Schweingruber}, R.~F., {Freiherr von Forstner},
  J.~L., {Wang}, Y., {Dresing}, N., et~al. (2020).
\newblock {First Solar Energetic Particles Measured on the Lunar Far-side}.
\newblock \emph{apjl} 902, L30.
\newblock \doi{10.3847/2041-8213/abbccc}
\bibAnnoteFile{Xu2020ApJ}

\bibitem[{Zhang et~al.(2020)Zhang, Wimmer-Schweingruber, Yu, Wang, Fu, Zou
  et~al.}]{zhang-etal-2020}
Zhang, S., Wimmer-Schweingruber, R.~F., Yu, J., Wang, C., Fu, Q., Zou, Y.,
  et~al. (2020).
\newblock First measurements of the radiation dose on the lunar surface.
\newblock \emph{Science Advances} 6.
\newblock \doi{10.1126/sciadv.aaz1334}
\bibAnnoteFile{zhang-etal-2020}

\end{thebibliography}






\section*{Supplementary material}

# Supplementary Material

## 1 CHANGES OF THE LND CONFIGURATION

Tab.S1 lists the changes to the LND configuration which were made on the 3rd, 4th, 5th, and 9th lunar day (counted from the landing of Chang'E 4). An unforeseen sudden drop in temperature caused mechanical stresses in detectors A, H, I, and J which unfortunately resulted in an increased noise level of these detectors. This necessitated changes 1 - 3 presented below. Furthermore, we updated the onboard software to improve the response to the highest-energy stopping protons as is discussed in bullet 4 beneath. Point 5 below discusses how the LND firmware accumulates data into DPS boxes. Because these had to be defined before LND could be fully calibrated, they are not placed optimally for heavy ions.

Table S1. L1 threshold Changes history of LND (upper) and other configuration changes (bottom).

| Lunar day | A2H | H2 | I1H | I2H | J1 |
|---|---|---|---|---|---|
| 3rd | 200 keV | 42 keV | 96 keV | 300 keV | 128 keV |
| 4th | 400 keV | | | | |
| 9th | | 300 keV | | | |
| 5th | Disable A2 in L3 logics: LET and charged particles | | Raise L3 threshold of detector I from 50 keV to 400 keV | | |

1. The stress-induced increase in the noise level of the affected detector segments were partially compensated by adjusting the thresholds of the affected segments, as shown in Tab. S1.

2. We disabled the A2 channel in the L3 logic of LET and stopping particles to further reduce the impact of the increased noise on the LET spectra. LND triggers on the B detector which is also used for determining the LET spectrum. If the signal in detector A is also above threshold, then the appropriate LET channel is augmented. Thus the increased noise in A2 resulted in a corrupted LET spectrum which was fixed by disabling the A2 channel in the L3 logic. Disabling A2 in the level-3 logic dramatically changed the geometry factors and path length of LET spectra. An updated version of Table 4 in (Wimmer-Schweingruber et al., 2020) regarding the parameter of various LET spectra is given in Tab. S2

3. In the fifth lunar day we uploaded commands to raise the threshold of detector I in L3 trigger logic from 50 keV to 400 keV. The data products of penetrating particle were changed mostly as shown in panel (b) of Fig. 2, since we filtered many minimum ionizing protons that deposited

| | Det. | # bins | $\langle L \rangle$ [$\mu$m] | $\hat{L}$[$\mu$m] | var($L$) [$\mu$m] | g [mm$^2$sr] |
|---|---|---|---|---|---|---|
| 1 | $A \cdot B \cdot C \cdot I$ | 64 | 523 | 523 | 54 | 33 |
| 2 | $\bar{A} \cdot B \cdot I \cdot J$ | 64 | 615 | 588 | 14896 | 2098 |
| 3 | $A \cdot B \cdot I \cdot J$ | $8 \times 64$ | 511 | 513 | 1178 | 79 |
| 4 | $2 \cup 3 \approx B \cdot J$ | 64 | 611 | 584 | 14774 | 2177 |
| 5 | $1 \cup 3 \approx A \cdot B$ | 64 | 514 | 515 | 881 | 112 |

Table S2. Properties of the various LET spectra after disabling A2 in LND. This is thus an updated version of Table 4 in Wimmer-Schweingruber et al. (2020)

energy less than 400 keV in detector I. These missing particles mainly populate in the left half of penetrating panels of Xmas plots in DPS box H12 and H10. Luckily the detector response of the albedo particles, which are also in the left half, did not change as its energy deposition in the front detectors are much higher than the level 3 threshold of detector I.

4. Protons stopping in the I detector deposit different energy in it than our initial simulations had predicted. This means that the H9 DPS box is not correctly placed and therefore only has a reduced geometry factor compared to the analytical value of 0.275 cm$^2$sr. The correct value is given in Tab. S3.

5. Because of an error in the calculation of the locations of the DPS boxes for heavy ions they do not reflect the real positions of heavy ions in the Xmas plot. Thus, the DPS heavy ions should not be used for analysis. Also, because of the very low fluxes of heavy ions, the better time resolution of the DPS boxes of heavy ions is not needed and the data provided by the Xmas plot (at 1 hour time resolution) is more than adequate. The positions of the DPS boxes have not been corrected.

Those changes in the LND configuration were successful in salvaging the primary data products of LND. The quality of LND's data continues to be excellent thanks to the redundant design of LND.

## 2 UPDATED RESPONSE FUNCTIONS FOR PROTONS

Figure 3 (a) shows the geometry factors for the different DPS channels. The lower and upper limits in energy of the proton DPS boxes are fairly sharp for stopping particles (DPS boxes H1 - H9), but are not as clearly delineated for penetrating particles, as is to be expected. In other words, the geometry factors are not purely geometric quantities, but energy dependent response functions. We give the updated boundaries at the 50%-level (a kind of FWHM) in the second column of Tab. S3. For the lowest energy bin (H1), we used the 10%-level for the lower boundary, for H10 we used the 10% boundary for the upper limit.

Because of the sometimes broad energy response (e.g., for H10 or H11) variations of the primary particle spectrum will result in different counts rates in these DPS channels even if the total number of particles remains the same. The count rate is determined by integrating the product of the geometry factor and the primary particle spectrum. This results in a small, but in some cases non-negligible variation of the geometry factor. This is shown in columns three and four of Tab. S3 for a logarithmically flat (power-law exponent $\gamma = -1$) and the solar-minimum 24/25 CREME96 model. The calculation was performed as shown in eq. S1,

$$\bar{G} = \frac{\int_0^{inf} f(E)\, g\, \mathrm{d}E}{\int_{E_{min}}^{E_{max}} f(E)\, \mathrm{d}E}, \tag{S1}$$

where $f(E)$ is the input spectrum and the $g$ are the (energy-dependent) geometry factors as shown in Fig. 3 (a).

Fig. 2 (a) shows that the proton DPS boxes are contaminated (primarily) by heavier ions. We account for this by a fixed correction factor for each DPS channel which is given in columns 5 and 6 for the CREME96 and BON14 models (again computed for solar minimum 24/25). These corrections can be substantial for some channels, as can be seen in the table.

The last column of Tab. S3 gives the numerical values of the LND data points shown in Fig. 5 together with their uncertainties.

Table S3. Weighted geometry factors and energy bins of primary and albedo protons, correction factors caused by other particles for CREME96 and BON14 case, and the averaged GCR and albedo proton fluxes.

| Name | Energy(MeV) | GF($cm^2 sr$) ($\gamma$ = -1) | GF($cm^2 sr$) (CREME) | Correction (CREME) | Correction (BON14) | GCR Flux $(cm^2\ sr\ s\ Mev)^{-1}$ |
|---|---|---|---|---|---|---|
| H1 | 9.18, 10.73 | 0.29 | 0.30 | 0.95 | 0.87 | $4.66 \pm 0.28 \times 10^{-5}$ |
| H2 | 10.73, 12.90 | 0.28 | 0.31 | 0.97 | 0.94 | $3.96 \pm 0.21 \times 10^{-5}$ |
| H3 | 12.90, 15.96 | 0.29 | 0.32 | 0.96 | 0.94 | $4.22 \pm 0.18 \times 10^{-5}$ |
| H4 | 15.96, 18.65 | 0.27 | 0.27 | 0.98 | 0.97 | $4.90 \pm 0.23 \times 10^{-5}$ |
| H5. | 18.65, 21.17 | 0.26 | 0.28 | 0.97 | 0.96 | $5.29 \pm 0.24 \times 10^{-5}$ |
| H6. | 21.17, 29.73 | 0.27 | 0.28 | 0.96 | 0.96 | $7.60 \pm 0.15 \times 10^{-5}$ |
| H7 | 29.73, 31.50 | 0.28 | 0.27 | 0.98 | 0.97 | $8.38 \pm 0.37 \times 10^{-5}$ |
| H8 | 31.50, 33.36 | 0.25 | 0.29 | 0.82 | 0.78 | $7.96 \pm 0.34 \times 10^{-5}$ |
| H9 | 33.17, 34.14 | 0.17 | 0.20 | 0.79 | 0.76 | $9.13 \pm 0.61 \times 10^{-5}$ |
| H10 | 139.2, 368.4 | 0.11 | 0.095 | 0.98 | 0.98 | $4.74 \pm 0.01 \times 10^{-4}$ |
| H11 | 42.3, 139.2 | 0.28 | 0.30 | 0.36 | 0.35 | $1.98 \pm 0.01 \times 10^{-4}$ |
| H14 | 34.94, 41.76 | 0.28 | 0.35 | 0.75 | 0.72 | $1.54 \pm 0.02 \times 10^{-4}$ |
| Albedo | 64.7, 76.7 | 0.18 | - | - | - | $1.12 \pm 0.09 \times 10^{-4}$ |

REFERENCES

Wimmer-Schweingruber, R. F., Yu, J., Böttcher, S. I., Zhang, S., Burmeister, S., Lohf, H., et al. (2020). The Lunar Lander Neutron and Dosimetry (LND) Experiment on Chang'E 4. Space Science Reviews 216, 104. doi:10.1007/s11214-020-00725-3



\end{document}